\documentclass[10pt]{article}
\usepackage{graphicx}

\def\Title#1{\begin{center} {\Large #1 } \end{center}}
\def\Author#1{\begin{center}{ \sc #1} \end{center}}
\def\Address#1{\begin{center}{ \it #1} \end{center}}

\newcommand\pubblock{\rightline{\begin{tabular}{l} Proceedings of the Fifth Annual LHCP\\ \pubnumber\\
         \pubdate  \end{tabular}}}

\newenvironment{Abstract}{\begin{quotation} \begin{center} 
             \large ABSTRACT \end{center}\bigskip 
      \begin{center}\begin{large}}{\end{large}\end{center} \end{quotation}}

\newenvironment{Presented}{\begin{quotation} \begin{center} 
             PRESENTED AT\end{center}\bigskip 
      \begin{center}\begin{large}}{\end{large}\end{center} \end{quotation}}

\def\beq{\begin{equation}}
\def\eeq#1{\label{#1}\end{equation}}
\def\eeqn{\end{equation}}

\def\beqa{\begin{eqnarray}}
\def\eeqa#1{\label{#1}\end{eqnarray}}
\def\eeqan{\end{eqnarray}}

\let\bar=\overbar

\def\Dslash{\not{\hbox{\kern-4pt $D$}}}
\def\dslash{\not{\hbox{\kern-2pt $\del$}}}

\def\msb{{\bar{\ssstyle M \kern -1pt S}}}

\usepackage{color}

\newcommand\pubnumber{ CMS CR-2017/207}
\newcommand\pubdate{\today}

\def\affiliation{
On behalf of the ATLAS and CMS collaborations, \\
Institut f\"ur Experimentelle Teilchenphysik \\
Karlsruher Institut f\"ur Technologie, 76131 Karlsruhe, Germany}

\begin{document}

% large size for the first page
\large
\begin{titlepage}
\pubblock

%% Change the title, name, abstract
%% Title 
\vfill
\Title{Single top $t$-channel in ATLAS and CMS}
\vfill

%  if you need to add the support use this, fill the \support definition above. 
%   \Author{ FIRSTNAME LASTNAME \support }
\Author{ Nils Faltermann  }
\Address{\affiliation}
\vfill
\begin{Abstract}

  The production of single top quarks allows to study the interplay of top quark physics and the electroweak sector of the standard model. Deviations from predictions can be a hint for physics beyond the standard model. The $t$-channel is the dominant production mode for single top quarks at the LHC. This talk presents the latest measurements from the ATLAS and CMS collaborations.
  
\end{Abstract}
\vfill

% DO NOT CHANGE 
\begin{Presented}
The Fifth Annual Conference\\
 on Large Hadron Collider Physics \\
Shanghai Jiao Tong University, Shanghai, China\\ 
May 15-20, 2017
\end{Presented}
\vfill
\end{titlepage}
\def\thefootnote{\fnsymbol{footnote}}
\setcounter{footnote}{0}
%

% normal size for the rest
\normalsize 

%% Your paper should be entered below. 

\section{Introduction}
\begin{figure}[htb]
\centering
\includegraphics[width=0.15\textwidth]{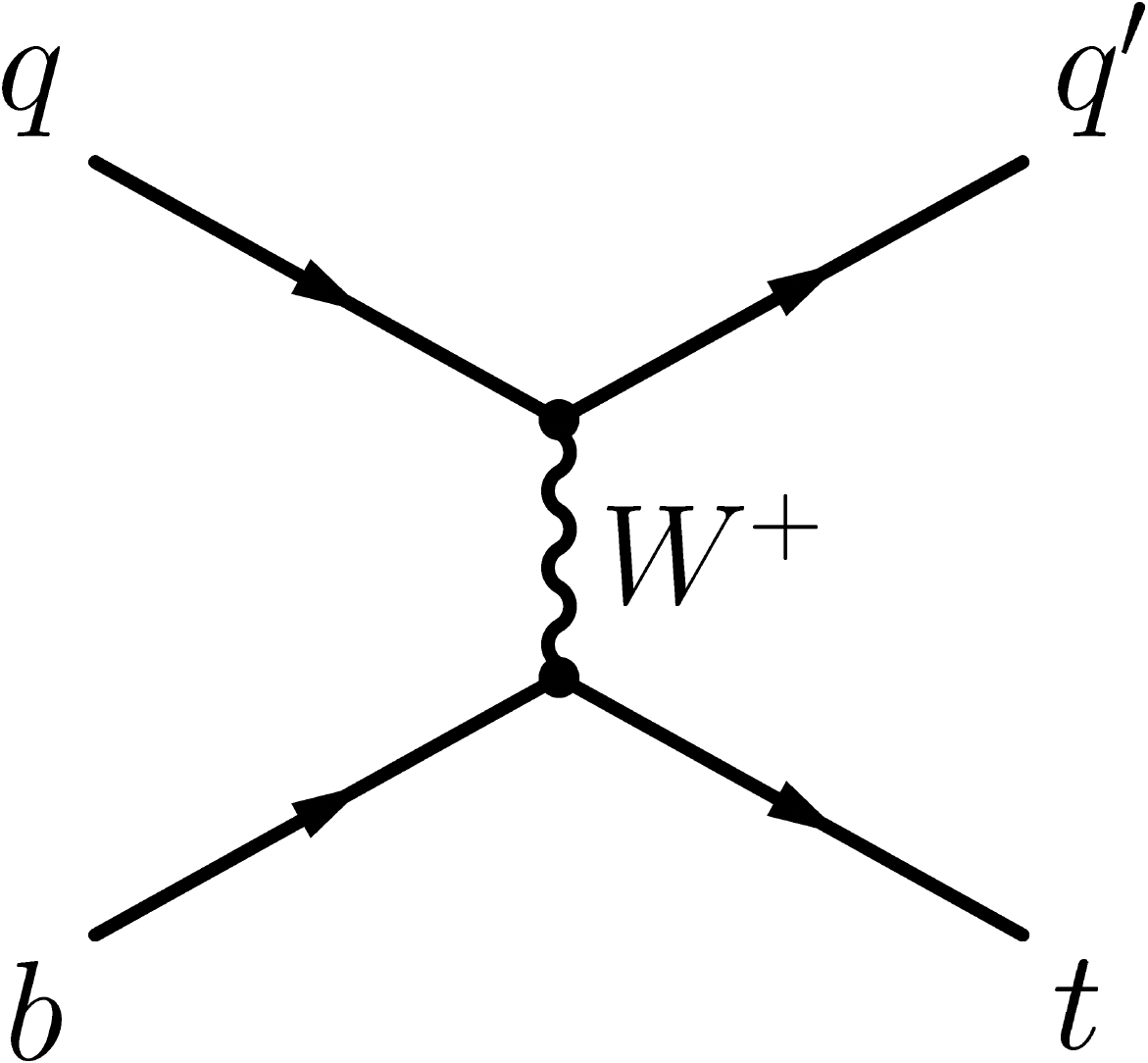}\hspace{1cm}
\includegraphics[width=0.15\textwidth]{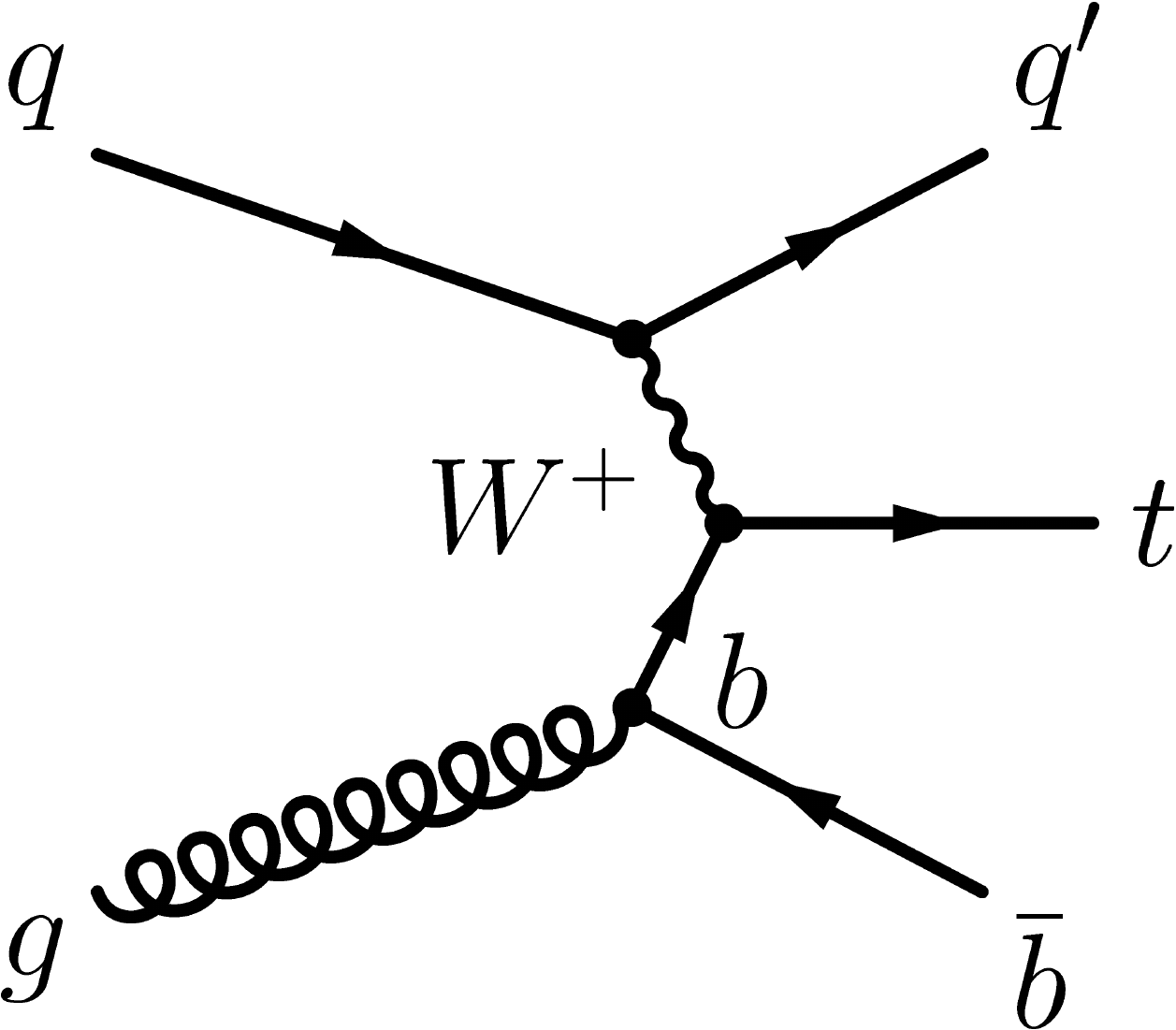}
\caption{Leading order Feynman diagrams for $t$-channel single top quark production at the LHC in the five-flavor scheme (left) and the four-flavor scheme (right).}
\label{fig:fdia}
\end{figure}

Unlike top quark pair production, which is induced via the strong interaction, single top quarks are produced through electroweak interaction. This allows to probe the electroweak sector of the standard model with precision measurements. Deviations in such observables may be a hint for physics beyond the standard model. The production of single top quarks allows to study the Wtb vertex directly, as well as to measure the $V_{\mathrm{tb}}$ matrix element of the CKM matrix.
\\
Single top quarks at the LHC are mostly produced via the $t$-channel. The other production modes, namely the $s$-channel and tW-associated production, only contribute to about 30\% to the total single top production rate. In the $t$-channel a top quark is produced through the exchange of a W boson between a light-flavored quark and a bottom quark, changing the flavor of the bottom quark to a top quark. The description of this process can be done in two different schemes, the five-flavor (5FS) scheme and the four-flavor (4FS) scheme. In the 5FS scheme the assumption is that the initial bottom quark comes directly from the proton, while in the 4FS scheme this bottom quark comes from an additional gluon splitting. This makes the 5FS scheme a 2$\rightarrow$2 process and the 4FS a 2$\rightarrow$3 process. The leading order Feynmmn diagrams for both schemes are shown in Figure~\ref{fig:fdia}.
The predicted cross section for top and antitop production in the $t$-channel at 8 TeV are:
\begin{eqnarray}
  \sigma_{t\mathrm{-ch.,t}}^{8\,\mathrm{TeV}}  &=& 54.87  ^{+1.64} _{-1.09} \, (\mathrm{scale}) \pm 1.60 \, (\mathrm{PDF}+\alpha_{\mathrm{S}}) \, \mathrm{pb},\\
    \sigma_{t\mathrm{-ch.,\bar{t}}}^{8\,\mathrm{TeV}}  &=& 29.74  ^{+0.92} _{-0.59} \, (\mathrm{scale}) \pm 1.39 \, (\mathrm{PDF}+\alpha_{\mathrm{S}}) \, \mathrm{pb},\\
    \sigma_{t\mathrm{-ch.,t+\bar{t}}}^{8\,\mathrm{TeV}}  &=& 84.69  ^{+2.56} _{-1.68} \, (\mathrm{scale}) \pm 2.76 \, (\mathrm{PDF}+\alpha_{\mathrm{S}}) \, \mathrm{pb}.
\end{eqnarray}
All values are obtained using the \textsc{HATHOR} program~\cite{Kant:2014oha} in the 5FS scheme. A top quark mass of $m_{\mathrm{top}}=172.5 \, \mathrm{GeV}$ is used. The predictions for 13 TeV are:
\begin{eqnarray}
  \sigma_{t\mathrm{-ch.,t}}^{13\,\mathrm{TeV}}  &=& 136.02   ^{+4.09} _{-2.92} \, (\mathrm{scale}) \pm 3.52 \, (\mathrm{PDF}+\alpha_{\mathrm{S}}) \, \mathrm{pb},\\
    \sigma_{t\mathrm{-ch.,\bar{t}}}^{13\,\mathrm{TeV}}  &=& 80.95  ^{+2.53} _{-1.71} \, (\mathrm{scale}) \pm 3.18 \, (\mathrm{PDF}+\alpha_{\mathrm{S}}) \, \mathrm{pb},\\
    \sigma_{t\mathrm{-ch.,t+\bar{t}}}^{13\,\mathrm{TeV}}  &=& 216.99  ^{+6.62} _{-4.64} \, (\mathrm{scale}) \pm 6.16 \, (\mathrm{PDF}+\alpha_{\mathrm{S}}) \, \mathrm{pb}.
\end{eqnarray}

\section{Inclusive, differential and fiducial cross section measurement from the ATLAS collaboration at $\sqrt{s} = 8 \, \mathrm{TeV}$}
\begin{figure}[htb]
\centering
\includegraphics[width=0.27\textwidth]{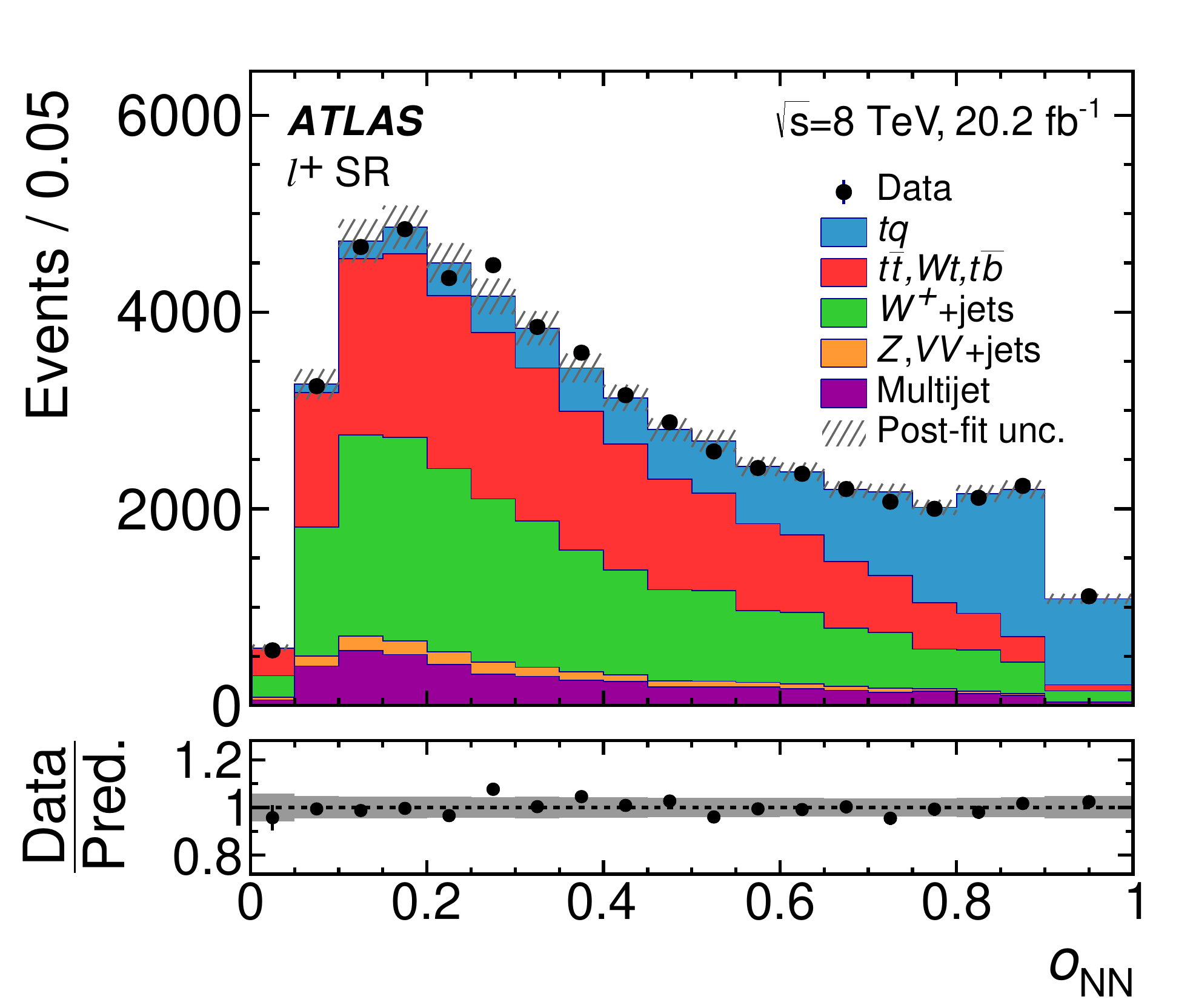}\hspace{1cm}
\includegraphics[width=0.27\textwidth]{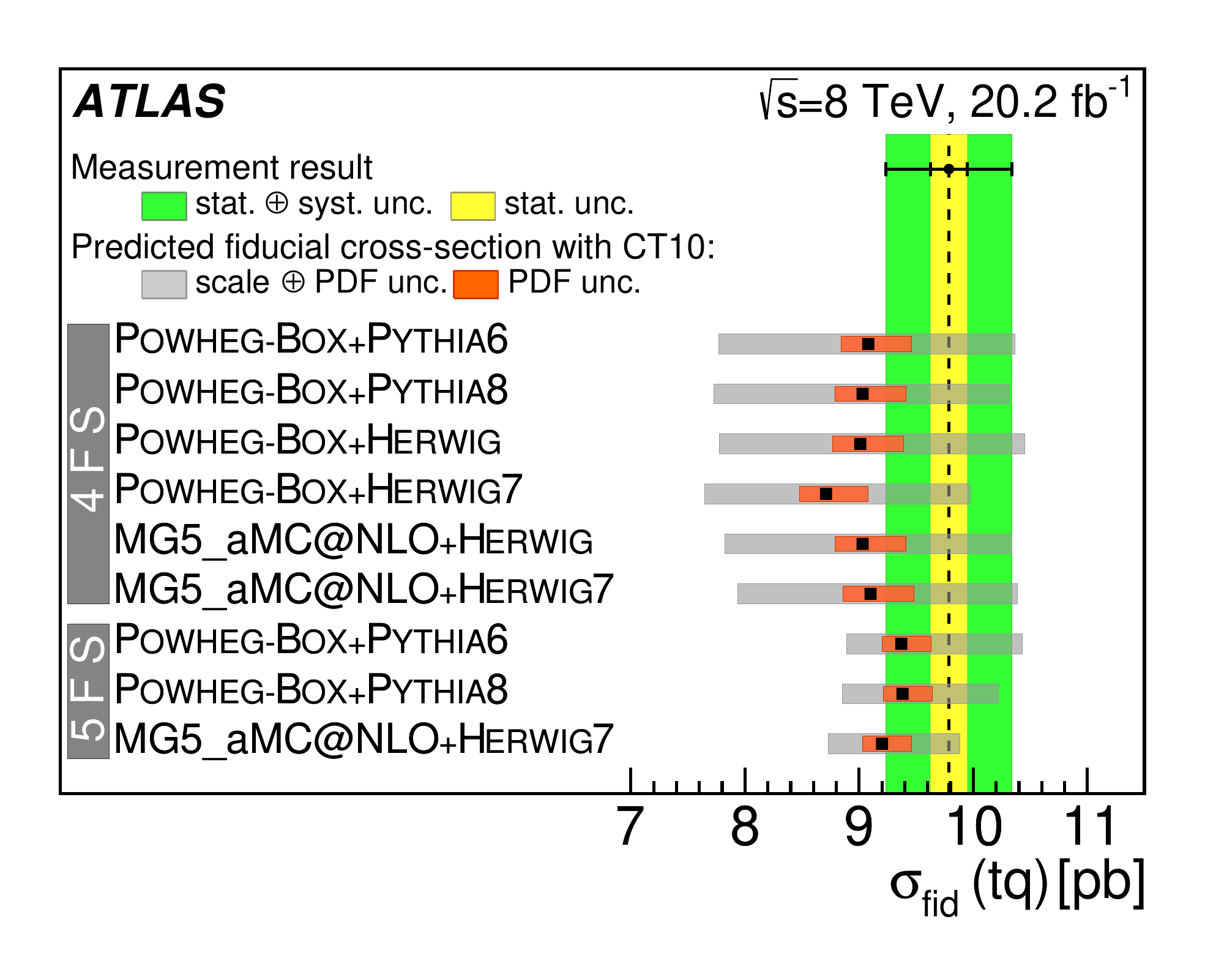}
\caption{Distribution of the neural network discriminant (left) and comparison of the measured fiducial cross section for top production in the $t$-channel with different predictions (right)~\cite{Aaboud:2017pdi}.}
\label{fig:atlas8_incl}
\end{figure}

\begin{figure}[htb]
\centering
\includegraphics[width=0.272\textwidth]{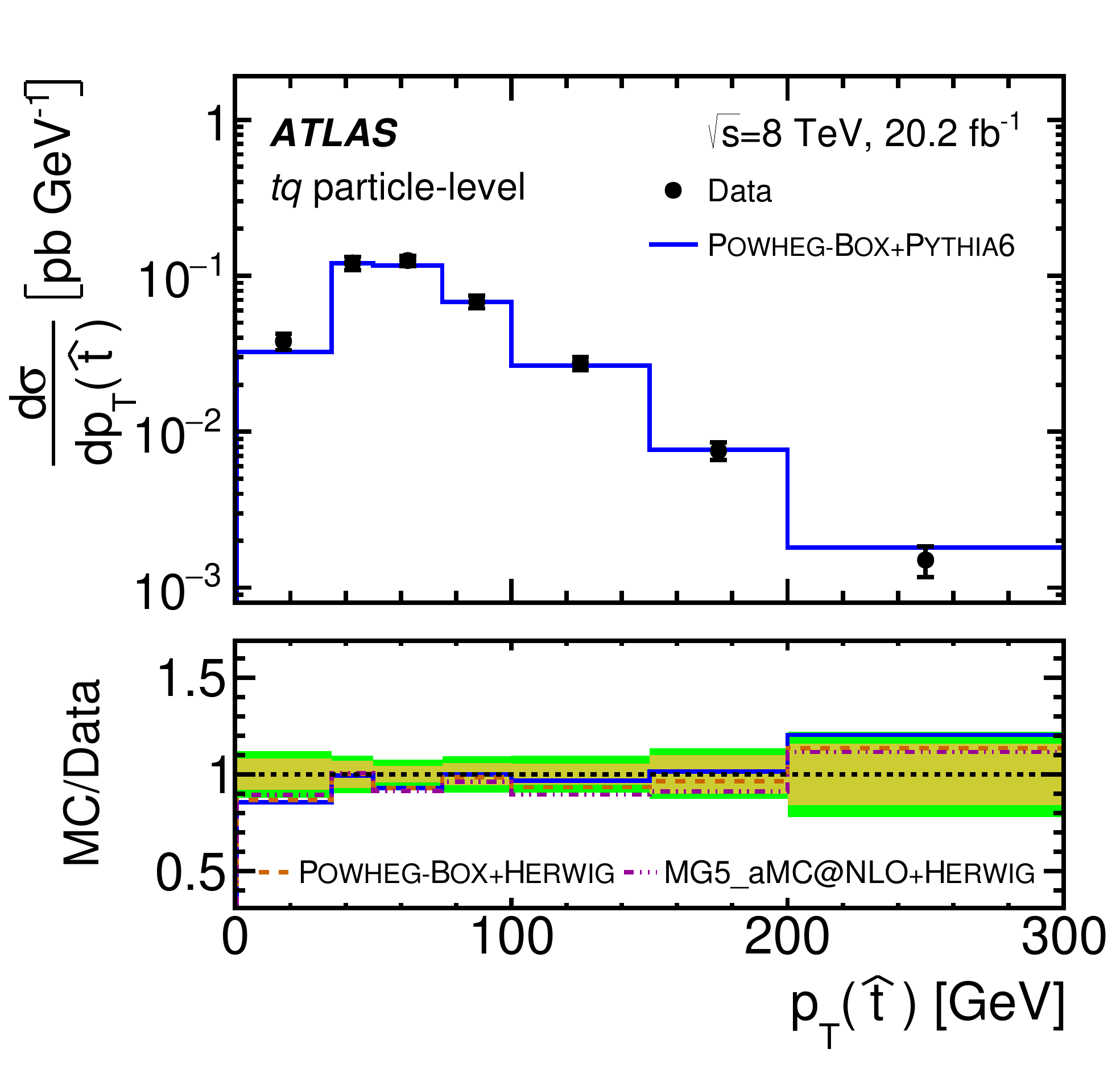}\hspace{1cm}
\includegraphics[width=0.272\textwidth]{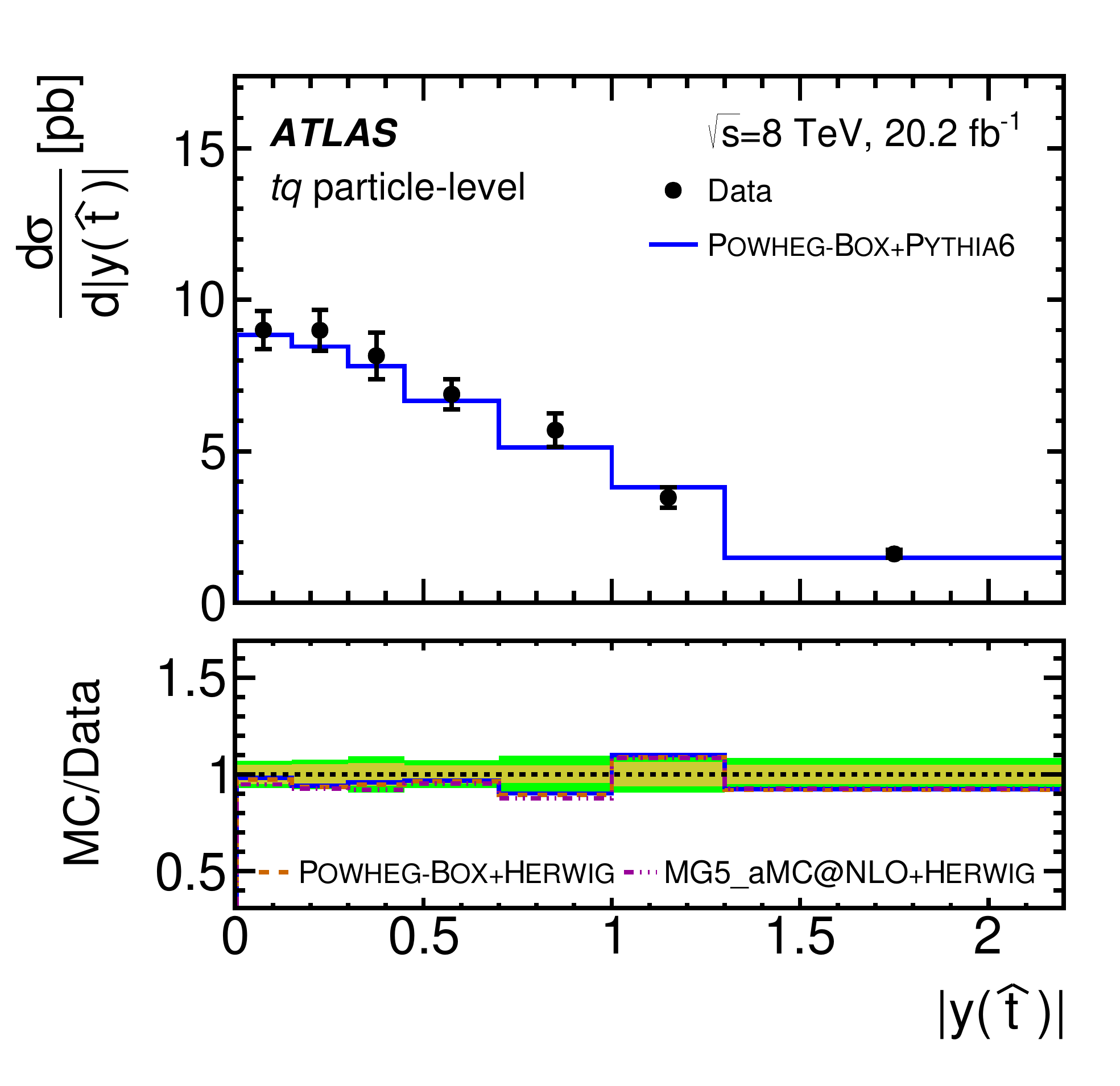}\\
\includegraphics[width=0.272\textwidth]{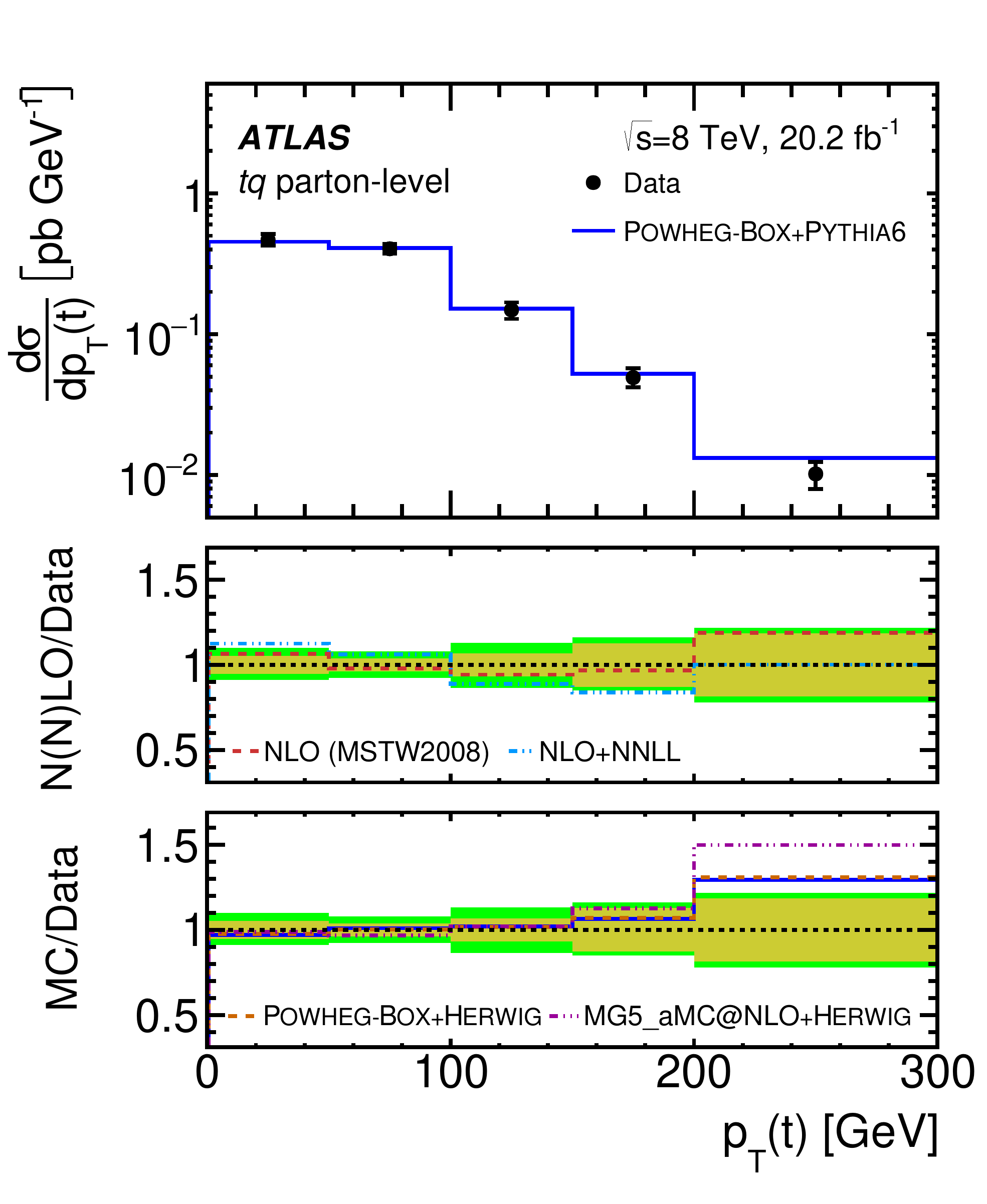}\hspace{1cm}
\includegraphics[width=0.272\textwidth]{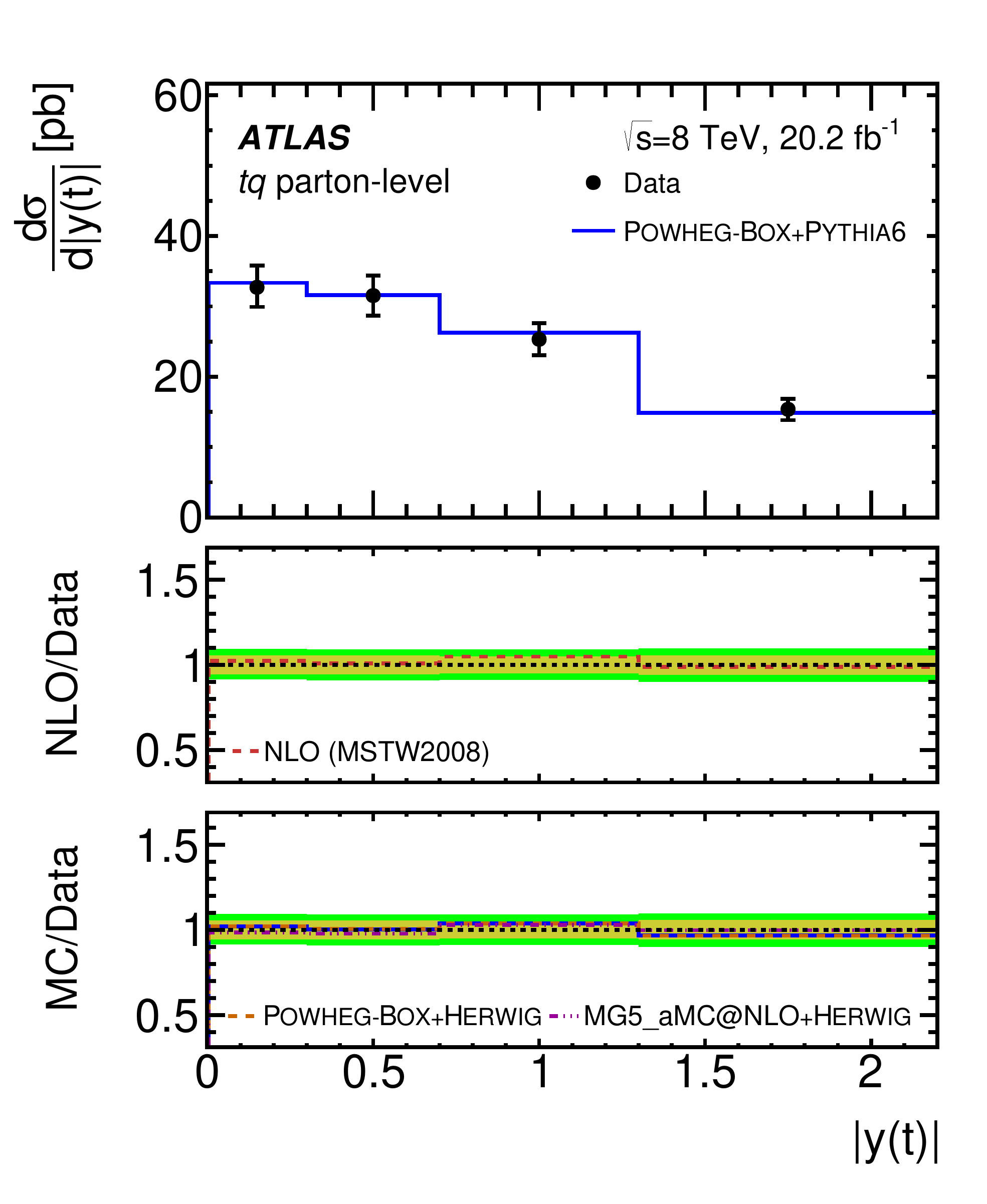}
\caption{Unfolded distributions for the transverse momentum (left) and the rapidity (right) of the top quark. The figures in the top row show the results for the particle-level definition, whereas the bottom row shows the results for the parton-level definition~\cite{Aaboud:2017pdi}.}
\label{fig:atlas8_diff}
\end{figure}

The analysis measures the cross section of top and antitop single top quark production at $\sqrt{s} = 8 \, \mathrm{TeV}$ with an integrated luminosity of $20.2~\mathrm{fb}^{-1}$. Details on the ATLAS experiment can be found in Ref.~\cite{Aad:2008zzm}. Both cross sections can be used to probe the parton density function (PDF) of the proton as the single top quark production is proportional to the up-type quark contribution of the PDF and the antitop quark production to the down-type quark contribution. Since many systematic uncertainties affect both cross sections in the same way it is advisable to define the ratio $R_t \equiv \sigma(tq) / \sigma(\bar{t}q)$ as an observable. The total cross section of single top quark production can be used to determine the CKM matrix element $V_{tb}$. In addition to the measurement in the full phase space, a fiducial measurement based on stable Monte Carlo particles is performed which is less affected by acceptance corrections.
\\
Only leptonic decays of the top quark are considered, i.e. the W boson originating from the Wtb vertex decaying into an electron or muon and a neutrino. The latter is not directly visible in the detector, but can be indirectly measured as missing transverse energy. Events are therefore selected to contain one electron or muon, a large fraction of missing transverse energy and two jets, where one of the two jets is identified as originating from a bottom quark.
\\
The main backgrounds contributing to the measurement are $W$+jets and top quark pair production ($t\bar{t}$). Predictions for signal and background processes are obtained from Monte Carlo simulations, except for the QCD multijet contribution which is obtained from a fit to the distribution of missing transverse energy. To improve the signal and background separation a neural network is trained which combines the separation power of different variables into one single discriminant. The inclusive cross section is then obtained with a maximum-likelihood fit to the distribution of the neural network output. The post-fit distribution and the comparison of the measured fiducial cross section for top quark production with different predictions are shown in Figure~\ref{fig:atlas8_incl}. The inclusive cross section for top and antitop quark production is found to be:
\begin{eqnarray}
  \sigma_{\mathrm{tot}}(tq)  &=& 56.7  ^{+4.3} _{-3.8} \, \mathrm{pb},\\
  \sigma_{\mathrm{tot}}(\bar{t}q)  &=& 32.9  ^{+3.0} _{-2.7}  \, \mathrm{pb}.
\end{eqnarray}
This results in a ratio of $R_t=1.72 \pm 0.09$. With the combined cross section a value of $1.029 \pm 0.048$ for $f_{\mathrm{LV}} \cdot |V_{tb}|$ is obtained, where $f_{\mathrm{LV}}$ is a left-handed form factor which can introduce contributions from new physics. Dominating systematic uncertainties are $t\bar{t}$ NLO matching, $t\bar{t}$ parton shower and jet energy scale. The latter is only relevant for the cross section measurement and cancels out in the ratio measurement.
\\
For the differential measurement a cut to the output of the classifier is applied to derive a signal-enriched region. The differential cross section of top and antitop is measured as a function of $p_\mathrm{T}$ and $|y|$ of the (anti)top quark. The measured distributions are unfolded to parton and particle level in order to compare them directly to theoretical predictions. The unfolded distributions are shown in Figure~\ref{fig:atlas8_diff}.
Dominating systematic uncertainties for the differential measurement are background normalization, $t\bar{t}$ NLO matching and signal QCD scale variations.
\\
All measured distributions and quantities are in agreement with standard model predictions. Additional information on the analysis can be found in Ref.~\cite{Aaboud:2017pdi}.

\section{Inclusive cross section measurement from the CMS collaboration at $\sqrt{s} = 13 \, \mathrm{TeV}$}
\label{cms_incl}
\begin{figure}[htb]
\centering
\includegraphics[width=0.28\textwidth]{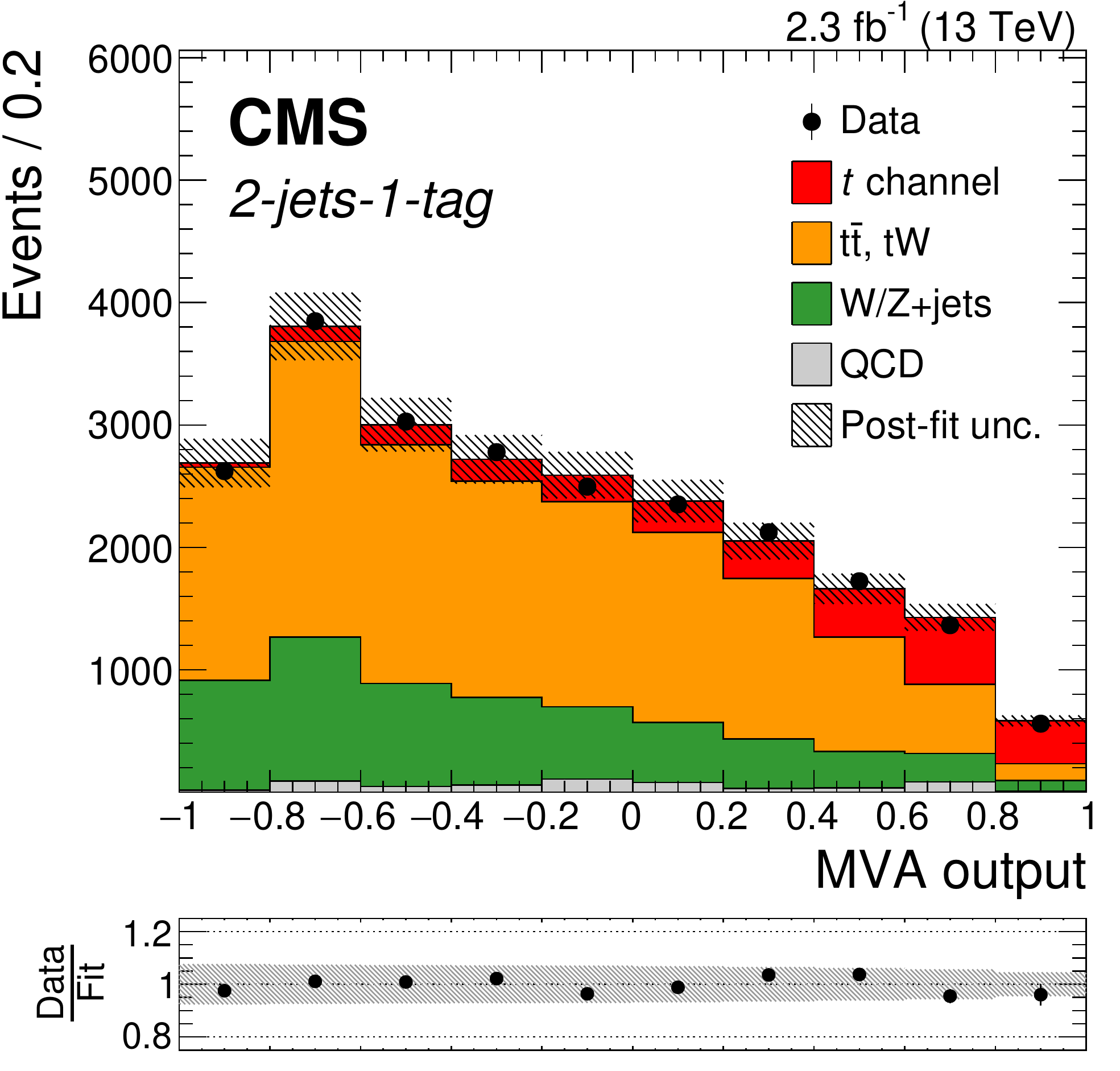}\hspace{1cm}
\includegraphics[width=0.38\textwidth]{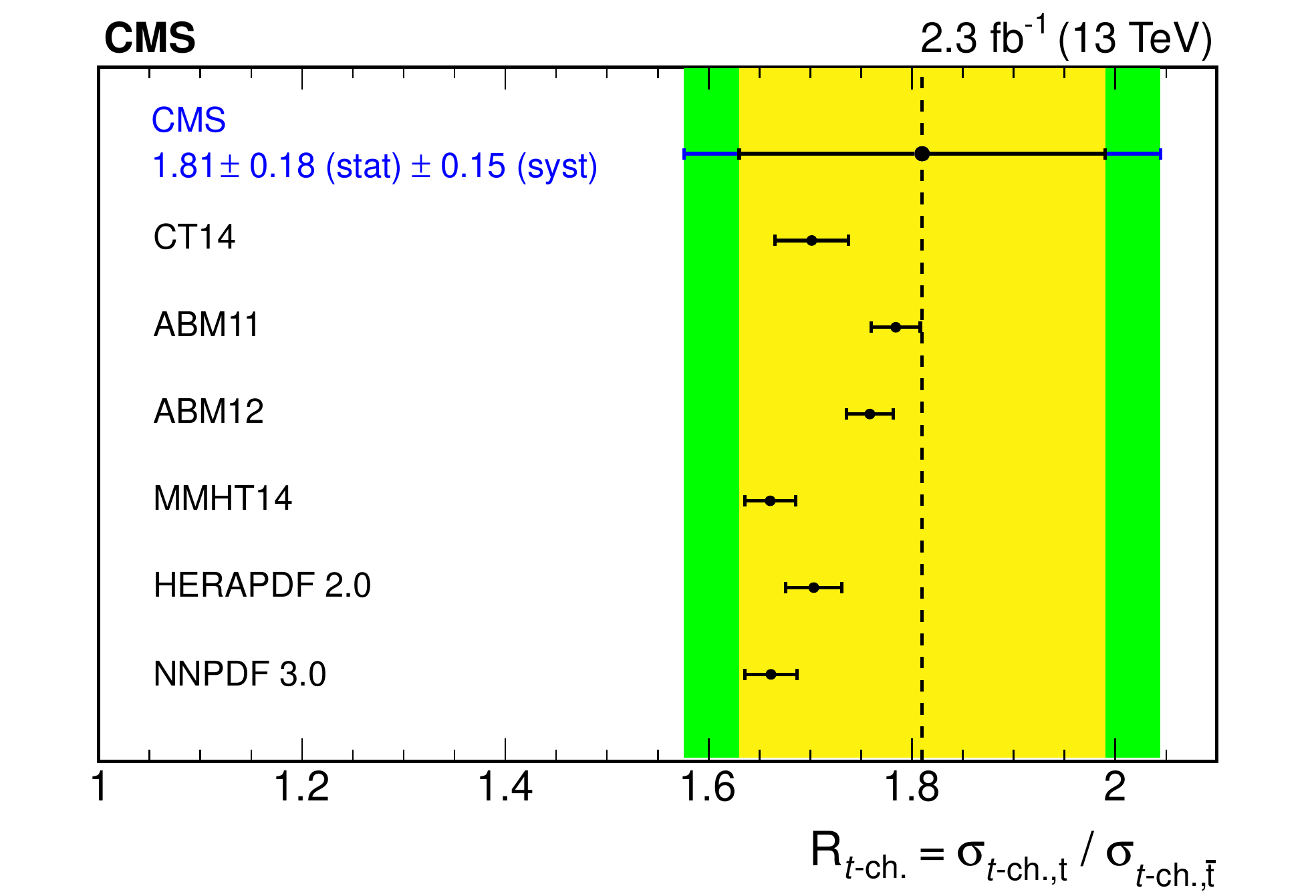}
\caption{Distribution of the neural network discriminant (left) and comparison of the measured top/antitop ratio with different predictions (right)~\cite{Sirunyan:2016cdg}.}
\label{fig:cms13_incl}
\end{figure}

In this analysis the inclusive cross section of single top quark production is measured at a center-of-mass energy of $\sqrt{s}~=~13 \, \mathrm{TeV}$. The analyzed data corresponds to an integrated luminosity of $2.3~\mathrm{fb}^{-1}$. A description of the CMS experiment can be found in Ref.~\cite{Chatrchyan:2008aa}.
\\
A signal-enriched region is defined which contains one isolated muon and two jets, where one of the two jets is identified as originating from a bottom quark. The background contribution from QCD multijet events is estimated with a data-driven approach. By inverting the muon isolation a QCD-enriched sideband with high purity is defined. This template is then used in a maximum-likelihood fit to the $m_\mathrm{T}$ distribution in the isolated region to derive the QCD contribution. To extract the signal cross section a maximum-likelihood fit is performed to the distribution of a neural network classifier. Additional control regions are used in the fit to constrain the contribution from $\mathrm{t}\mathrm{\bar{t}}$ background events.
\\
Figure~\ref{fig:cms13_incl} shows the post-fit distribution of the neural network output and the comparison of the measured ratio with predictions from different PDF sets. The measured cross sections are:
\begin{eqnarray}
  \sigma_{t\mathrm{-ch.,t}}  &=& 150 \pm 8 \mathrm{(stat)} \pm 9 \mathrm{(exp)} \pm 18 \mathrm{(theo)} \pm 4 \mathrm{(lumi)} \, \mathrm{pb} , \\
  \sigma_{t\mathrm{-ch.,\bar{t}}}  &=& 82 \pm 10 \mathrm{(stat)} \pm 4 \mathrm{(exp)} \pm 11 \mathrm{(theo)} \pm 2 \mathrm{(lumi)} \, \mathrm{pb}.
\end{eqnarray}
The ratio of top and antitop quark production is found to be $R_{\mathrm{t-ch.}} = 1.81 \pm 0.18\mathrm{(stat)} \pm 0.15 \mathrm{(syst)}$ and the total cross section results in $f_{\mathrm{LV}} \cdot |V_{\mathrm{tb}}| = 1.03 \pm 0.07 \mathrm{(exp)} \pm 0.02 \mathrm{(theo)}$. Systematic uncertainties with the largest impact on the measurement are signal modeling, QCD scale variations and jet energy scale. No significant deviation from the standard model predictions is observed. More details can be found in Ref.~\cite{Sirunyan:2016cdg}.

\section{Inclusive cross section measurement from the ATLAS collaboration at $\sqrt{s} = 13 \, \mathrm{TeV}$}
\begin{figure}[htb]
\centering
\includegraphics[width=0.28\textwidth]{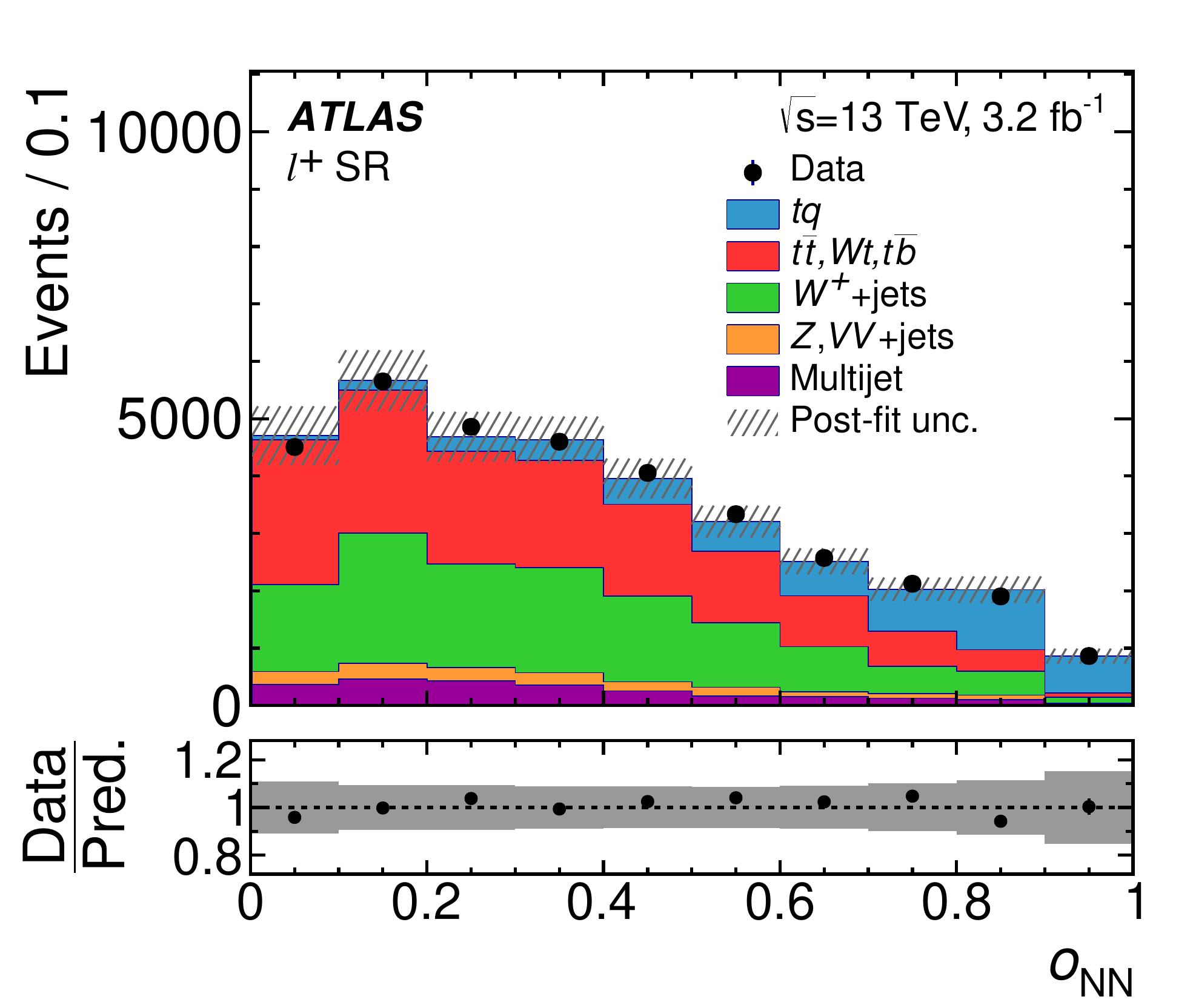}\hspace{1cm}
\includegraphics[width=0.38\textwidth]{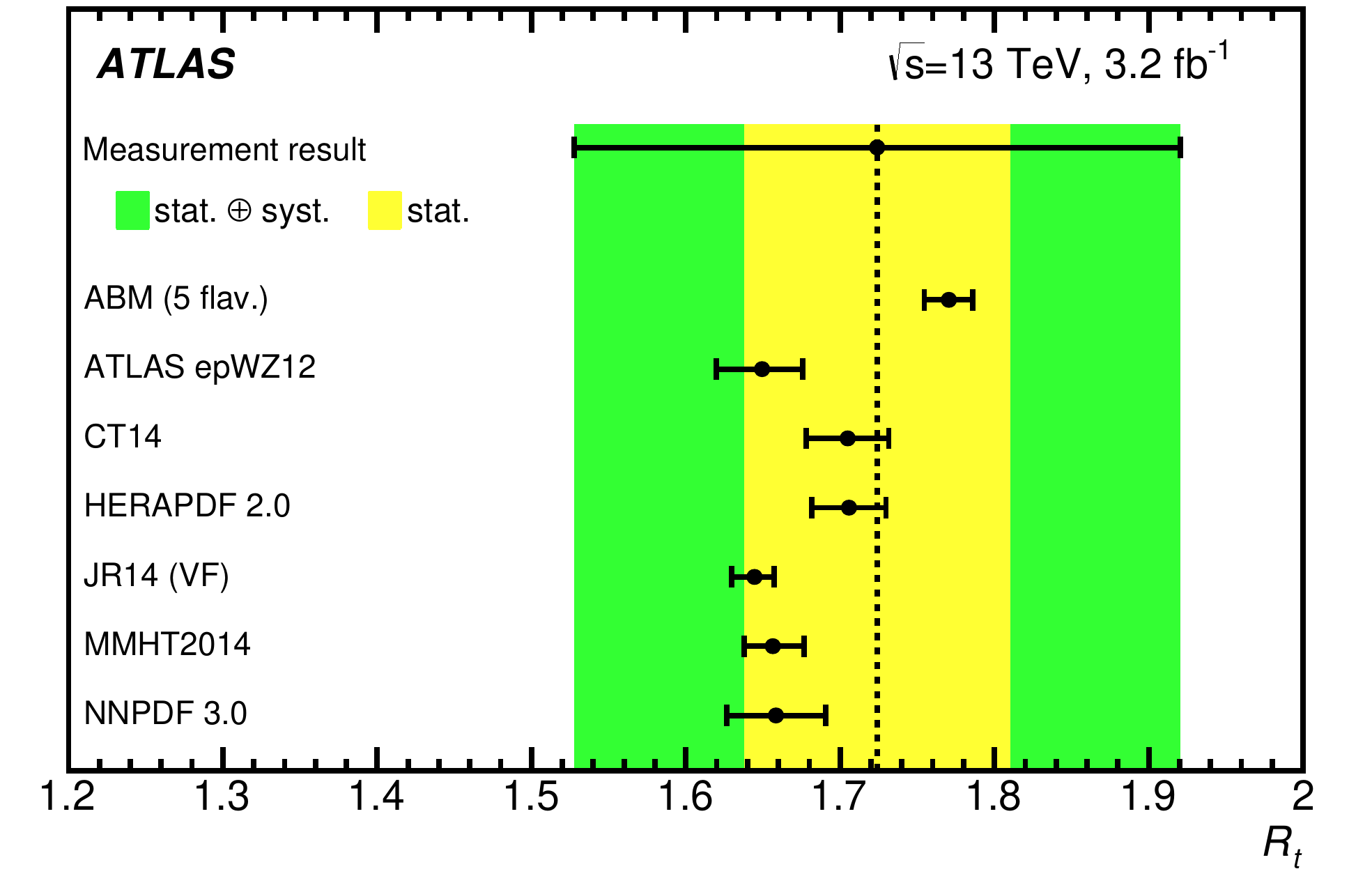}
\caption{Distribution of the neural network discriminant (left) and comparison of the measured top/antitop ratio with different predictions (right)~\cite{Aaboud:2016ymp}.}
\label{fig:atlas13_incl}
\end{figure}

The inclusive cross section at $\sqrt{s}~=~13 \, \mathrm{TeV}$ is also measured by the ATLAS collaboration with an integrated luminosity of $3.2~\mathrm{fb}^{-1}$.
\\
Events are selected which contain one electron or muon and two jets, where one jet is identified as b jet. Two additional regions are defined to validate the modeling of $W$+jets and $t\bar{t}$ background. A neural network is trained to classify events as signal- or background-like. The cross section is then extracted with a maximum-likelihood fit to the output distribution of this classifier.
\\
The post-fit distribution of the neural network classifier output is shown in Figure~\ref{fig:atlas13_incl}, as well as the comparison of the measured ratio with different PDF sets. The following cross sections are measured:
\begin{eqnarray}
  \sigma (tq)  &=& 156 \pm 5 \mathrm{(stat)} \pm 27 \mathrm{(syst)} \pm 3 \mathrm{(lumi)} \, \mathrm{pb},\\
  \sigma (\bar{t}q)  &=& 91 \pm 4 \mathrm{(stat)} \pm 18 \mathrm{(syst)} \pm 2 \mathrm{(lumi)} \, \mathrm{pb}.
\end{eqnarray}
The ratio is determined to be $R_{t} = 1.72 \pm 0.09\mathrm{(stat)} \pm 0.18 \mathrm{(syst)}$ and the total cross section results in $f_{\mathrm{LV}} \cdot |V_{tb}| = 1.07 \pm 0.01 \mathrm{(stat)} \pm 0.09 \mathrm{(syst)} \pm 0.02 \mathrm{(theo)} \pm 0.01 \mathrm{(lumi)}$. The most dominating uncertainties are b tagging efficiency, signal parton shower and background NLO matching. No deviations from the standard model predictions are observed. The analysis is described in detail in Ref.~\cite{Aaboud:2016ymp}.

\section{Differential cross section measurement from the CMS collaboration at $\sqrt{s} = 13 \, \mathrm{TeV}$}
\begin{figure}[htb]
\centering
\includegraphics[width=0.3\textwidth]{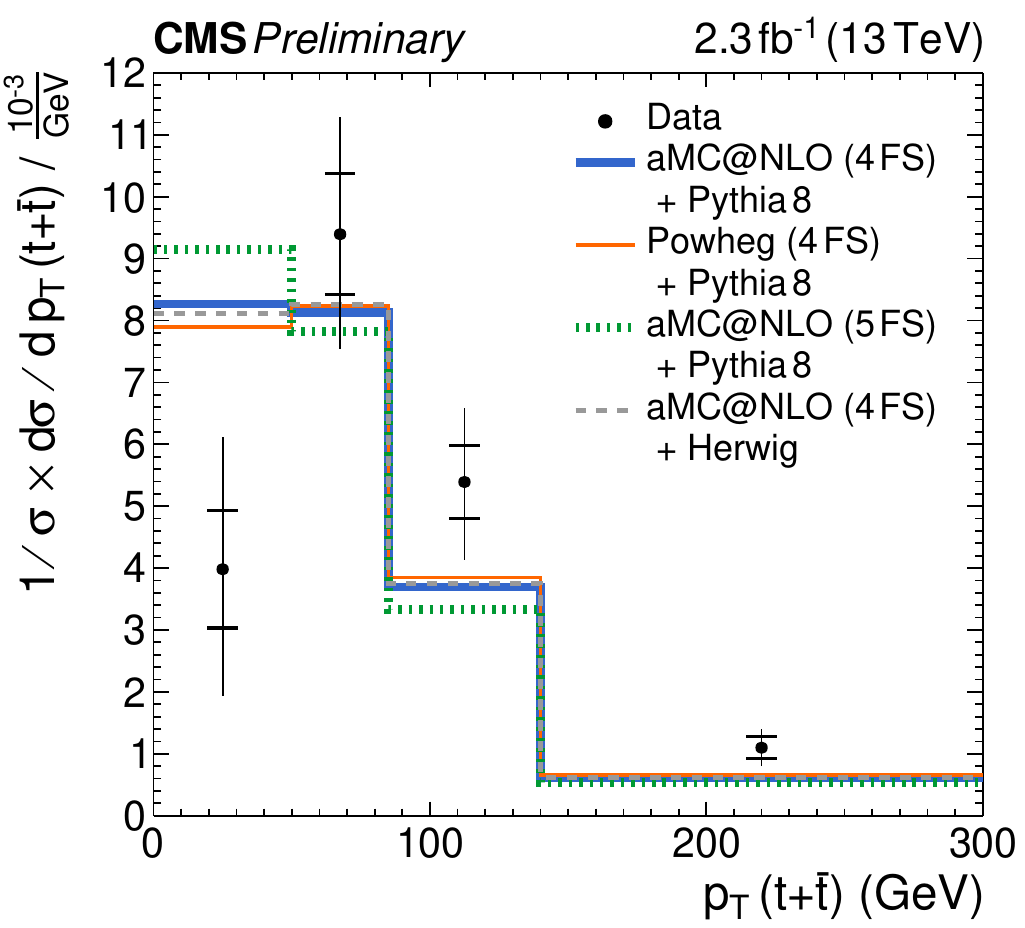}\hspace{1cm}
\includegraphics[width=0.3\textwidth]{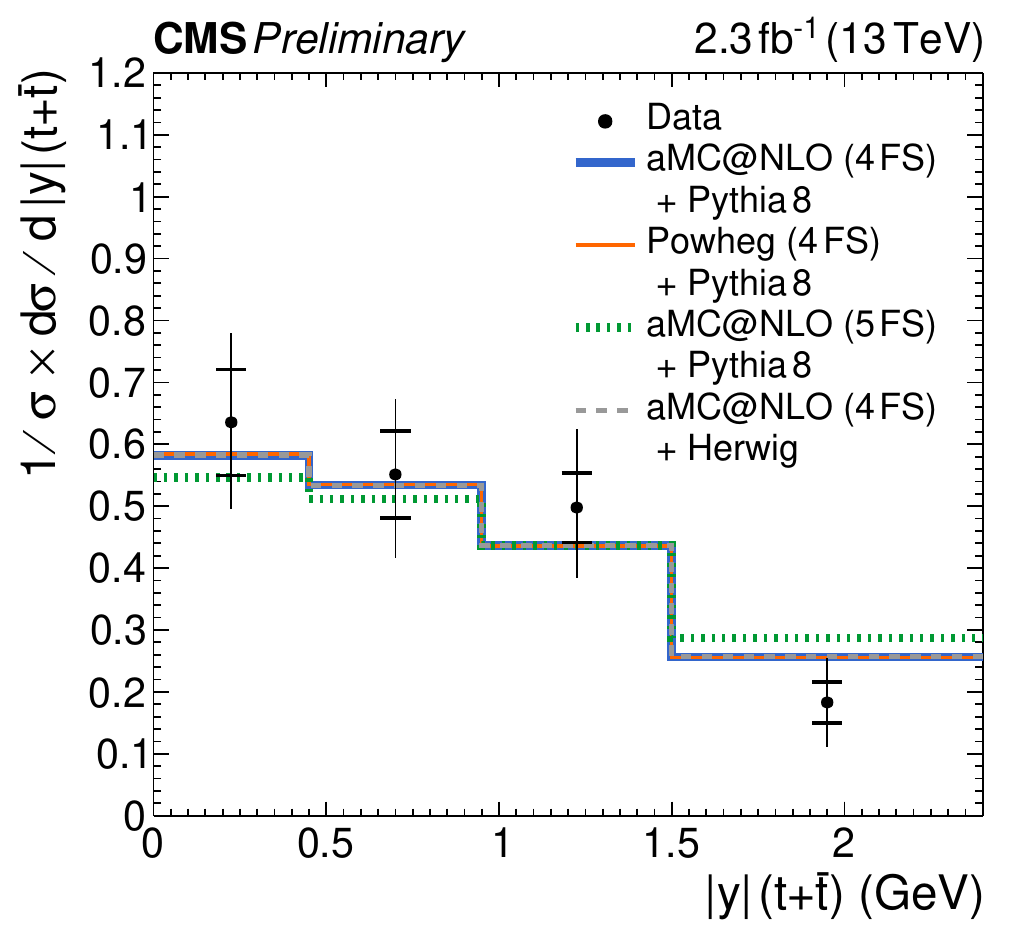}
\caption{Unfolded distributions for the transverse momentum (left) and the rapidity (right) of the top and antitop quark~\cite{CMS:2016xnv}.}
\label{fig:cms13_diff}
\end{figure}

A first measurement of the differential cross section of single top quark production at $\sqrt{s}~=~13 \, \mathrm{TeV}$ is performed by the CMS collaboration. The analyzed data corresponds to an integrated luminosity of $2.3~\mathrm{fb}^{-1}$.
\\
The selection is the same as in the inclusive measurement by the CMS collaboration in Section~\ref{cms_incl}. Signal and background events are classified with the help of a boosted decision tree. A combined maximum-likelihood fit is performed to the output distribution of the classifier and the $m_\mathrm{T}$ distribution, which is used to estimate the QCD background contribution.
\\
The differential cross section is measured as a function of $p_\mathrm{T}$ and $|y|$ of the top and antitop quark. Both distributions are unfolded to parton level to compare them directly with theoretical predictions. The unfolded distributions are shown in Figure~\ref{fig:cms13_diff} together with predictions from different Monte Carlo simulations. Uncertainties with the largest impact on the measurement are the data statistics, the QCD scale of the signal, the top quark mass and the jet energy scale. Additional information can be found in Ref.~\cite{CMS:2016xnv}.

\section{Conclusion}
Both the ATLAS and CMS collaborations have measured the single top quark cross section in detail. The inclusive results are used to probe the proton PDF and to measure the CKM matrix element $V_{\mathrm{tb}}$, while differential distributions allow to precisely test simulations. All measured quantities and distributions are in agreement with standard model predictions.

%%%%%%%%%%%%%%%%%%%%%%%%%%%%%%%%%%%%%%%%%%%%%%%%%%%%%%%%%%%%%%%%%%%%%%%%%
%%
%%   use this format to include an .eps figure into your paper
%%
%% \begin{figure}[htb]
%% \centering
%% \includegraphics[height=2in]{head_lhcp2017.jpg}
%% \caption{ Place the caption here}
%% \label{fig:figure1}
%% \end{figure}
%%%%%%%%%%%%%%%%%%%%%%%%%%%%%%%%%%%%%%%%%%%%%%%%%%%%%%%%%%%%%%%%%%%%%%%%%%%

%% See Figure \ref{fig:figure1} and Table \ref{tab:table1}. 

%%%%%%%%%%%%%%%%%%%%%%%%%%%%%%%%%%%%%%%%%%%%%%%%%%%%%%%%%%%%%%%%%%%%%%%%%
%%
%%   use this format to include a LaTeX table  into your paper
%%
%% \begin{table}[t]
%% \begin{center}
%% \begin{tabular}{l|ccc}  
%% Patient &  Initial level($\mu$g/cc) &  w. Magnet &  
%% w. Magnet and Sound \\ \hline
%%  Guglielmo B.  &   0.12     &     0.10      &     0.001  \\
%%  Ferrando di N. &  0.15     &     0.11      &  $< 0.0005$ \\ \hline
%% \end{tabular}
%% \caption{ place the caption here }
%% \label{tab:table1}
%% \end{center}
%% \end{table}
%%%%%%%%%%%%%%%%%%%%%%%%%%%%%%%%%%%%%%%%%%%%%%%%%%%%%%%%%%%%%%%%%%%%%%%%%%%

%%  if necessary
%% \Acknowledgements
%% I am grateful to XYZ for fruitful discussions.

\end{document}